# Flexible over-moded resonators based on P(VDF-TrFE) thin films with very high temperature coefficient

V. S. Nguyen[1], L. Badie[2], E. Sénéchault[3], E. Blampain[1], B. Vincent[1], C. Venet[3], O. Elmazria[1], D. Rouxel[1]

[1] Institut Jean Lamour, UMR CNRS 7198, Faculté des Sciences et Techniques, Université de Lorraine, 54506, Vandœuvre-lès-Nancy, France

[2] Structure et Réactivité des Systèmes Moléculaires Complexes, UMR CNRS 7565, Faculté des Sciences et Techniques, Université de Lorraine, 54506, Vandœuvre-lès-Nancy, France

[3] Schneider Electric, 38TEC, T3, 37 Quai Paul Louis Merlin, 38050, Grenoble, France

**Abstract**

This work presents for the first time a flexible over-moded resonators based on P(VDF-TrFE) thin films. The devices were manufactured on commercial elastic substrate with inkjet printed electrodes. The sensing copolymer films used in the devices were polarized by corona method after electrode deposition. The main performance parameters of the component were then determined. The performed over mode resonator (OMR) on P(VDF-TrFE) exhibited a linear variation of frequency versus temperature and a very large value of temperature coefficient of frequency (TCF > 1600 ppm/°C). These properties suggest a great potential for using such components as low cost and high precision temperature sensors. The electromechanical coupling coefficient and the quality factor of the resonator were also characterized versus temperature.

**Introduction**

In the last decade, there has been great interest in applying polymers for microelectromechanical systems (MEMS) [1, 2], electronic memories [3] and further, in developing all-polymer systems [4]. There are more and more applications in which polymers play the role as functional components, entering sometimes in competition with crystalline materials (such as $LiNbO_3$ or $ZnO$ [5, 6]) for some acoustic devices. For the desired electroactive functionalities, among the organic materials, the copolymer of vinylidene difluoride and trifluoroethylene (P(VDF-TrFE)) is one of the best candidate thanks to its high piezoelectric activity without mechanical stretching, light weight, ease in processing, chemical and electrical resistance, and flexibility. The applications of this polymer included pyro- and piezo-electric sensors, pressure sensors, transducer, hydrophones…[7-9]. Furthermore, this kind of polymer is biocompatible [10], making it very attractive for biological applications [11, 12].

Thin-film bulk acoustic resonators (FBAR) are nowadays widely used in modern wireless systems, including radio frequency filters, duplexers, oscillators [13]. FBAR devices can also be used as sensing elements in applications ranging from pressure, chemical sensors to biological/DNA sensors [14-16]. It consists of a piezoelectric thin membrane sandwiched between two electrodes. When the thin piezoelectric transducers are attached to substrates and generated a standing wave resonance in the substrate (with thickness of multiple of the half acoustic wavelengths) [17], the devices are called over-moded resonators (OMR).

Today, printed electronic offers a promising way of fabricating electrical devices with attractive features such as low cost, large volume and sustainable production on flexible substrates [18]. It was predicted that the compound annual growth rate of printed and flexible sensors will be 22% over the next ten years [19]. Printed and flexible sensors are creating new markets using their unique advantages of flexibility, area and functionality [19-21].

In this work, we demonstrated the flexible OMR based on P(VDF-TrFE) thin films with inkjet printed electrodes. Some important performance parameters of the resonator were also characterized.

**Experimental**

Copolymer of vinylidene difluoride and trifluoroethylene (P(VDF-TrFE)) (70/30 mol %) in form of pellets was supplied by Piezotech S.A.S. (Hésingue, France). The detailed preparation and characterization of the copolymer films were described elsewhere [22, 23]. Briefly, the copolymer was dissolved in Methyl Ethyl Ketone (MEK, from Acros Organics) at 80 °C under stirring.

To fabricate the devices, a commercial flexible sheet of polyethylene terephthalate (PET) with thickness about 120 μm was used as substrate. First, the bottom electrode composed of a 100-nm thick gold layer and a 5-nm thick adhesion layer of chromium was deposed by evaporation on the PET. Then a 6.3-μm thick P(VDF-TrFE) film was spin-coated onto the metal electrode. Finally, the top electrode was printed by inkjet system (fig. 1). This last operation consisted in cleaning the surface to be printed with ethanol and deionized water before an atmospheric plasma treatment. The plasma treatment was a punctual plasma jet in atmospheric environment with the moving velocity of 40 mm/s and the distance to the sample of 3.5 cm. The pattern was then drawn (fig.1b) at 254 dpi using an inkjet printer and a silver ink containing 20

Wt.% of silver nanoparticles. The components were annealed at 140°C to obtain high piezoelectric phase crystallinity and hence high activity [22-24].

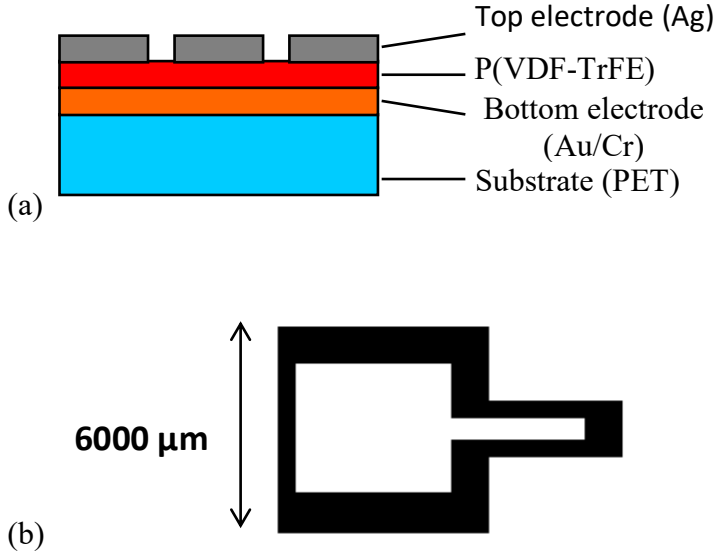

(a)

(b)

Figure 1: (a) cross-section and (b) design of the top electrode of OMR

To get high piezoelectric properties, the devices were polarized by corona discharge using a home-made setup with a voltage of 14 kV for 5 min, the grid - sample distance was 5 mm. The resonators were then characterized by a network analyzer (N5230A, Agilent Technologies). The temperature coefficient of frequency (*TCF*) was calculated using the equation:

$$TCF = \frac{1}{f_0}\frac{df}{dT} \qquad \text{(Equation 1)}$$

Where $f_0$ is the operating frequency of the device at room temperature and *T* is temperature.

The electromechanical coupling coefficient ($k^2$) and quality factor (*Q*) were determined form the frequencies of anti-resonance ($F_a$) and resonance ($F_r$), given by [25]

$$K^2 = \frac{\pi}{2}\frac{F_r}{F_a}\tan\left(\frac{\pi}{2}\frac{F_a-F_r}{F_a}\right) \qquad \text{(Equation 2)}$$

$$Q_i = \left(\frac{F_i}{F_a-F_r}\right), \quad i = r \text{ or } a \qquad \text{(Equation 3)}$$

**Results and discussion**

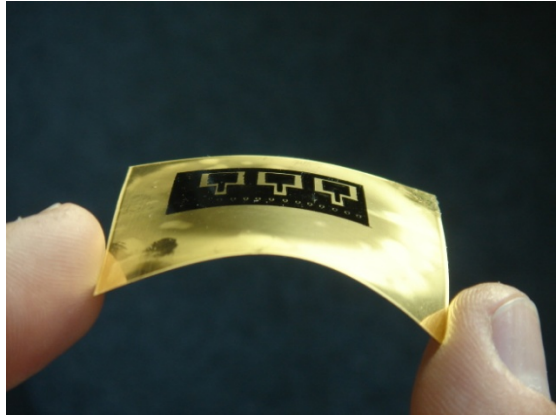

Figure 2: Photograph of the printed over-moded resonators

The photograph of the OMR of P(VDF-TrFE) on PET and the frequency response were showed in figure 2 and 3. Several resonant frequencies corresponding to different modes were observed. The resonance confirmed the good piezoelectric behavior of the copolymer films. As can be seen in figure 3, several narrow peaks could be used as operating frequency for the device. Moreover, their very close intensity and inaccuracies in measurement (e.g. calibrations) made it difficult to distinguish in this case the best useful mode. That is why for our measures we considered four interesting frequencies in the range of 35-70 MHz.

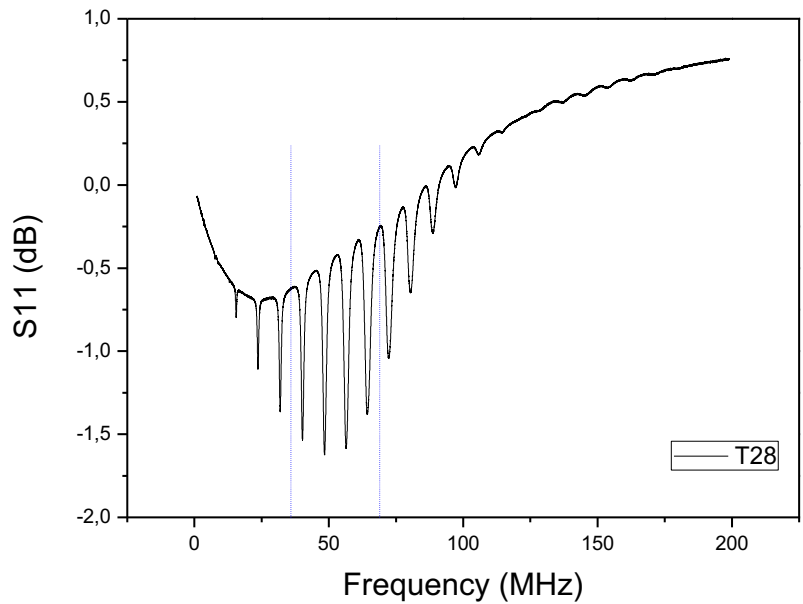

Figure 3: S11 spectrum of OMR of P(VDF-TrFE) on PET

The evolution of the reflection spectrum as a function of temperature was showed in figure 4. The resonance frequency strongly decreased with increasing temperature up to 90°C, attesting the high temperature sensitivity of the device. However, the signals gradually degraded and almost lost at 100 °C. This temperature was very closed to the Curie point (~103°C) of the P(VDF-TrFE) at which the transition ferro-paraelectric occurs.

The temperature coefficient of frequency (TCF) of the device was determined from the evolution of the operating frequency as a function of temperature (fig. 5). For this purpose, we have chosen to follow the frequency of the first minimum in figure 3. To our knowledge, it is the first time that the TCF is measured for bulk acoustic devices on fluoropolymer films. We show that P(VDF-TrFE) based device exhibited a very high value of TCF (-1605 ppm/°C) which was extremely high compared to other piezoelectric materials such as lithium niobate (around -80 ppm/°C depending on considered orientation), aluminum nitride (-25 ppm/°C) or zinc oxide (-60 ppm/°C) [17]. Furthermore, the operating frequency exhibited a linear behavior with temperature which is very interesting for sensor applications. The results suggested a remarkable ability of using this device as very high sensitive temperature sensors. Sensitivity combined with material flexibility and low cost could be a decisive advantage against other conventionally used materials.

The limitation lied in the working temperature range, because the copolymer cannot rise above 100 °C for not approaching the Curie temperature, and thus become depolarized and inactive. But thanks to its biocompatibility, this kind of material is very promising for biosensor applications (in-vivo or in-vitro) where operating temperature is much less than 100 °C.

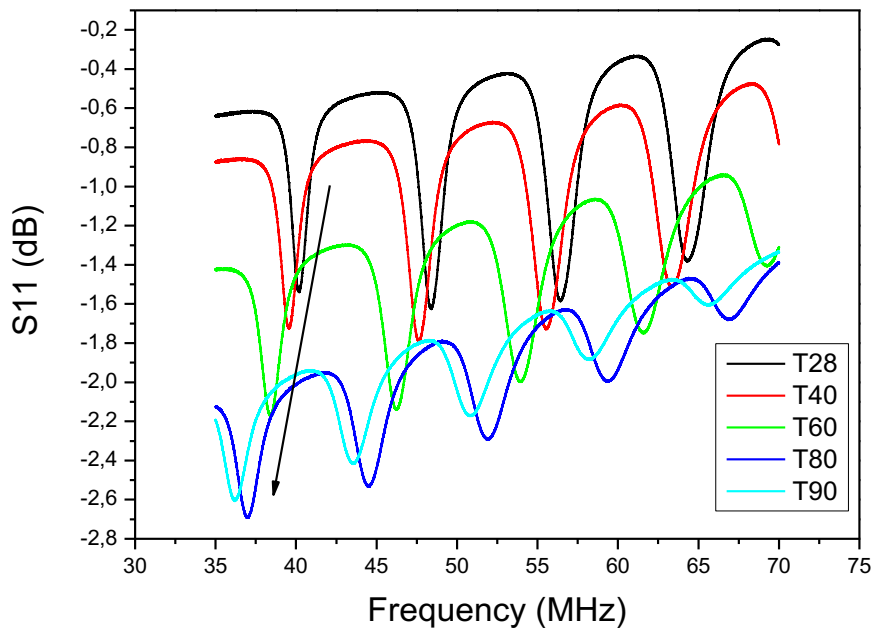

Figure 4: Evolution of S11 frequency responses as a function of temperature for OMR of P(VDF-TrFE) on PET

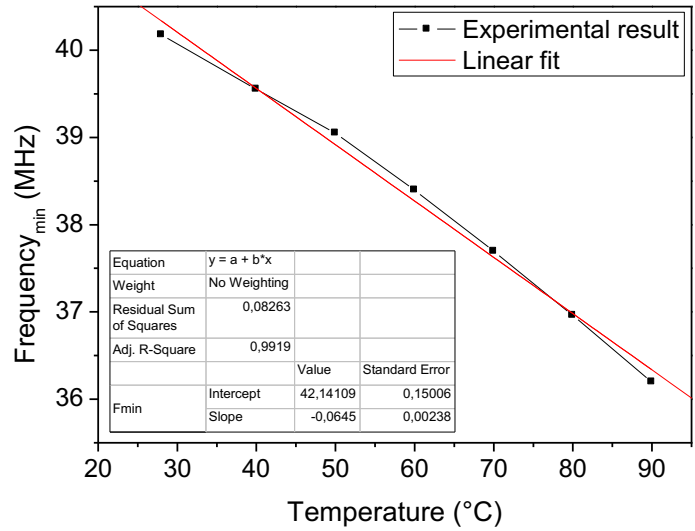

Figure 5: Evolution of the operating frequency as a function of temperature for OMR of P(VDF-TrFE) on PET

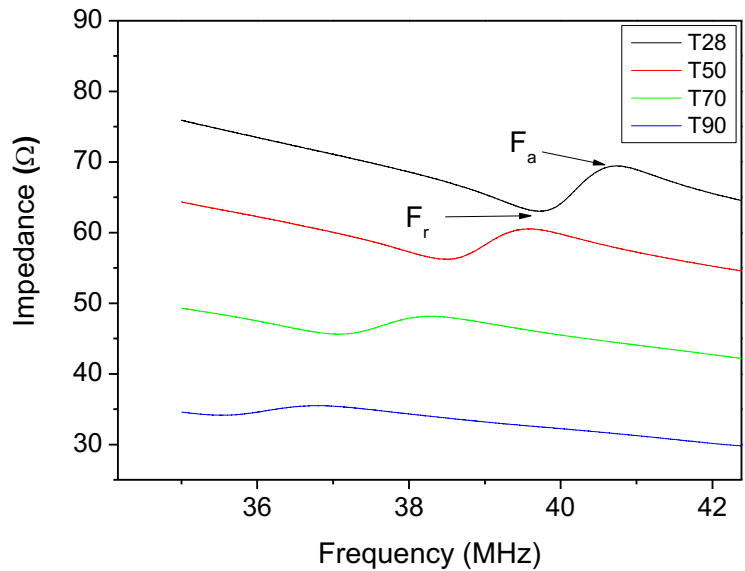

Figure 6: Impedance spectra as a function of temperature for OMR of P(VDF-TrFE) on PET

From the impedance spectra (fig. 6), we have determined two important parameters for the design of BAW components: the electromechanical coupling coefficient ($k^2$) and the quality factor (Q). At room temperature (28 °C), the $k^2$ of P(VDF-TrFE) resonators was quite large, by 5.9 %. This value is close to those provided the most coupled conventional piezoelectric materials such as aluminum nitride (6%), zinc oxide (7.5%) [17], Lithium Tantalate (5%), Lithium niobate (5-11%). The characterization versus temperature shows that this coefficient was enhanced when temperature increased. $k^2$ increased almost linearly and reached a value of 8.5% at 90 °C (fig. 7).

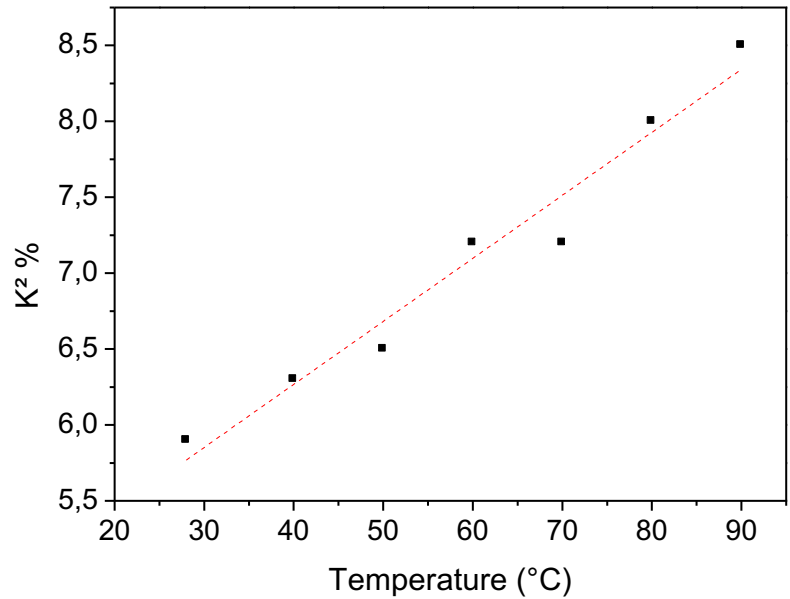

Figure 7: Evolution of the electromechanical coupling coefficient as a function of temperature for OMR of P(VDF-TrFE) on PET

The quality factor (Q) extracted from impedance-frequency curves using equation 3 is plotted in figure 8 versus temperature. Of course, the measured Q values were very low but these value could be improved using other configurations than OMR (i.e: FBAR, SMR [13, 25]). One can also observe that the Q factor decreased when temperature increased. However, this decrease of Q value was still in reasonable proportion if we consider the temperature range required for biological applications.

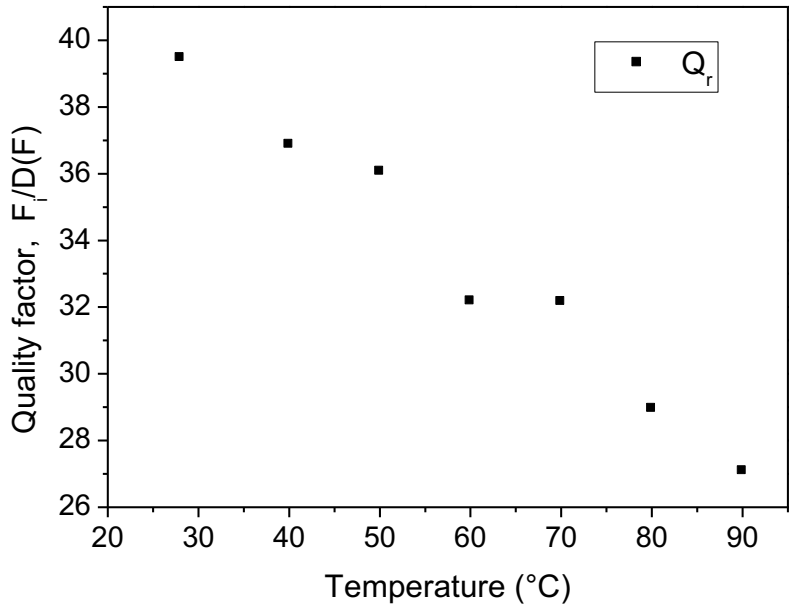

Figure 8: Evolution of the quality factor Q as a function of temperature for OMR of P(VDF-TrFE) on PET

**Conclusion**

Flexible over-moded resonators have been successfully prepared on P(VDF-TrFE) thin films. The devices were made on commercial PET substrates. The electrodes were printed by inkjet printing on spin-coating films of the copolymers and the devices were polarized by the non-contact corona discharge method to get high piezoelectric activity. This offers the prospect of low-cost routes of fabricating large-area flexible devices. Further, we found that the OMR of P(VDF-TrFE) exhibited a very high TCF which is very promising to realize low cost, high performance and high accuracy temperature sensors, especially for biological applications. In addition, the copolymer resonators had also quite good electromechanical coupling coefficient but moderate quality factor that could be enhanced by improving the resonator design.